\documentclass[preprint,12pt, a4paper]{elsarticle}

\usepackage{epsfig}

\usepackage{amssymb}

\usepackage{lineno}

\usepackage{hyperref}

\usepackage{float}
\restylefloat{table}

\journal{SoftwareX}

\begin{document}

\begin{frontmatter}

\title{Omicron: a tool to characterize transient noise in gravitational-wave detectors}

\author[ijclab]{Florent Robinet\corref{cor1}}
\author[ijclab]{Nicolas Arnaud}
\author[ijclab]{Nicolas Leroy}
\author[portsmouth]{Andrew Lundgren}
\author[cardiff]{Duncan Macleod}
\author[columbia]{Jessica McIver}
\cortext[cor1]{Corresponding author: \href{mailto:robinet@lal.in2p3.fr}{florent.robinet@ijclab.in2p3.fr} (F. Robinet).}

\address[ijclab]{Universit\'e Paris-Saclay, CNRS/IN2P3, IJCLab, 91405 Orsay, France}
\address[portsmouth]{Institute of Cosmology \& Gravitation, University of Portsmouth, Portsmouth PO1 3FX, UK}
\address[cardiff]{Cardiff University, Cardiff CF24 3AA, United Kingdom}
\address[columbia]{University of British Columbia, Vancouver V6T 1Z4, Canada }

\begin{abstract}
The Omicron software is a tool developed to perform a multi-resolution time--frequency analysis of data from gravitational-wave detectors: the LIGO, Virgo, and KAGRA detectors. Omicron generates spectrograms from whitened data streams, offering a visual representation of transient detector noises and gravitational-wave events. In addition, these events can be parameterized with an optimized resolution. They can be written to disk to conduct offline noise characterization and gravitational-wave event validation studies. Omicron is optimized to process, in parallel, thousands of data streams recorded by gravitational-wave detectors. The Omicron software plays an important role in vetting gravitational-wave detection candidates and characterization of transient noise.
\end{abstract}


\end{frontmatter}








\section{Motivation and significance}\label{sec:motivation}

In 2016, the LIGO~\cite{TheLIGOScientific:2014jea} and Virgo~\cite{Accadia:2015pda} collaborations announced the first detection of gravitational waves~\cite{Abbott:2016blz}. The signal detected by the LIGO interferometers, and labeled GW150914, originated from a system of two black holes spiraling down and merging into a single and final black hole. This event is of very short duration, $\sim200$~ms, and had to be extracted from a very large population of transient noise events, also called glitches, polluting the data of interferometric detectors.

Understanding glitches is critical to confidently claim the astrophysical nature of a gravitational-wave candidate. This noise investigation work relies on a monitoring of the instrument and its environment. For that purpose, tens of thousands of auxiliary data streams are recorded. They include data from environmental sensors (thermal, acoustic, seismic, magnetic...), photodiodes, actuators, electronic devices or feedback control loops. Many of those auxiliary channels are insensitive to gravitational waves and are used to witness disturbances from noise sources. More specifically, transient events must be searched for in thousands of auxiliary channels in order to identify correlations with the detector's output data and to understand coupling mechanisms leading to glitches. In this context, the Omicron algorithm, implementing a fast $Q$ transform~\cite{Brown:1991}, was developed to detect and characterize transient events in data from gravitational-wave detectors~\cite{Nuttall:2015dqa,TheLIGOScientific:2016zmo}.

Before 2011, at the time of the first generation of LIGO and Virgo detectors~\cite{Abbott:2007kv,Acernese:2008zza,Grote:2008zz}, the $Q$ transform was already implemented to search for transient gravitational-wave signals~\cite{Chatterji:2004,Abadie:2010mt,Abbott:2009zi} and for detector characterization work~\cite{Aasi:2012wd,Robinet:2010zz,Christensen:2010zz} but the \copyright{MATLAB} implementation was too slow to perform a multi-channel analysis required for detector characterization. Another approach, based on a wavelet decomposition~\cite{Chatterji:2004qg}, was much faster but offered poor precision on the parameter reconstruction. In this context, the Omicron algorithm was first developed in 2012 and much improved over the years. The code is written in C++ for performance reasons and lives in the LIGO--Virgo git repository~\cite{omicron-git}. A detailed description of the algorithm and its implementation can be found in the technical note~\cite{omicron-technote}.


\section{Software description}\label{sec:description}

The Omicron software is built upon GWOLLUM~\cite{GWOLLUM}, a set of libraries used to perform every analysis step of Omicron processing. The GWOLLUM package depends on several external libraries. The FrameL~\cite{FrameL} library is used to perform IO operations on data frame files~\footnote{The frame format was specifically developed for interferometric gravitational-wave detector data.}. Mathematical routines developed in the GNU Scientific Library~\cite{GSL} are commonly used by GWOLLUM functions. Discrete Fourier transforms are performed using the FFTW~\cite{FFTW} algorithm. Finally, the GWOLLUM and Omicron codes deeply rely on C++ classes developed in the CERN ROOT~\cite{Brun:1997pa} framework. For example, ROOT classes are used for plotting purposes, event storage or data access. To build Omicron and to manage dependencies, the CMake~\cite{cmake,cmake-paper} configuration tool was chosen.

Omicron~\cite{omicron-technote} is designed to search for power excess in data time series using the $Q$ transform~\cite{Brown:1991}. The $Q$ transform is a modification of the standard short Fourier transform in which the analysis window duration, $\sigma_t$, varies inversely with frequency, $\phi$, and is parameterized by a quality factor $Q$: $\sigma_t=Q/\sqrt{8}\pi\phi$. The data are projected onto a basis of complex-valued sinusoidal Gaussian functions, offering an optimal time--frequency decomposition of the signal $x$:
\begin{equation}
  X(\tau, \phi, Q) = \int_{-\infty}^{+\infty}{ x(t) w(t-\tau,\phi,Q) e^{-2i\pi\phi t}dt}.
  \label{eq:qtransform}
\end{equation}
The transform coefficient, $X$, measures the average signal amplitude and phase within a time--frequency region, called a \textit{tile}, centered on time $\tau$ and frequency $\phi$, whose shape and area are determined by the requested quality factor $Q$ and the Gaussian window, $w$, with a width $\sigma_t$. The parameter space $(\tau, \phi, Q)$ is tiled following the strategy developed in~\cite{Chatterji:2004}, where the tiles are distributed over a cubic lattice using a mismatch metric to guarantee a high detection efficiency. This tiling technique offers a multi-resolution time--frequency analysis that consists of logarithmically-spaced Q planes, logarithmically-spaced frequency rows, and linearly-spaced tiles in time.

Before applying the $Q$ transform, the input time series, $x(t)$, is normalized (or whitened) to $x^{wh}(t)$ such that, for pure stationary noise, the expectation value of the power spectral density is frequency-independent and is equal to~2. This whitening step offers a statistical interpretation for the $Q$ transform: the resulting transform coefficient, $X^{wh}$, is used to derive a signal-to-noise ratio estimator,
\begin{equation}
  \hat{\rho}(\tau,\phi,Q) =
  \left\{
  \begin{array}{ll}
    \sqrt{\left|X^{wh}(\tau, \phi, Q)\right|^2-2}, & \mathrm{if}\ \left|X^{wh}(\tau, \phi, Q)\right|^2 \ge 2\\
    0, & \mathrm{otherwise}.
  \end{array}
  \right.
  \label{eq:snr}
\end{equation}

\subsection{Software Architecture}\label{sec:architecture}

\begin{figure*}
  \center
  \includegraphics[width=12cm]{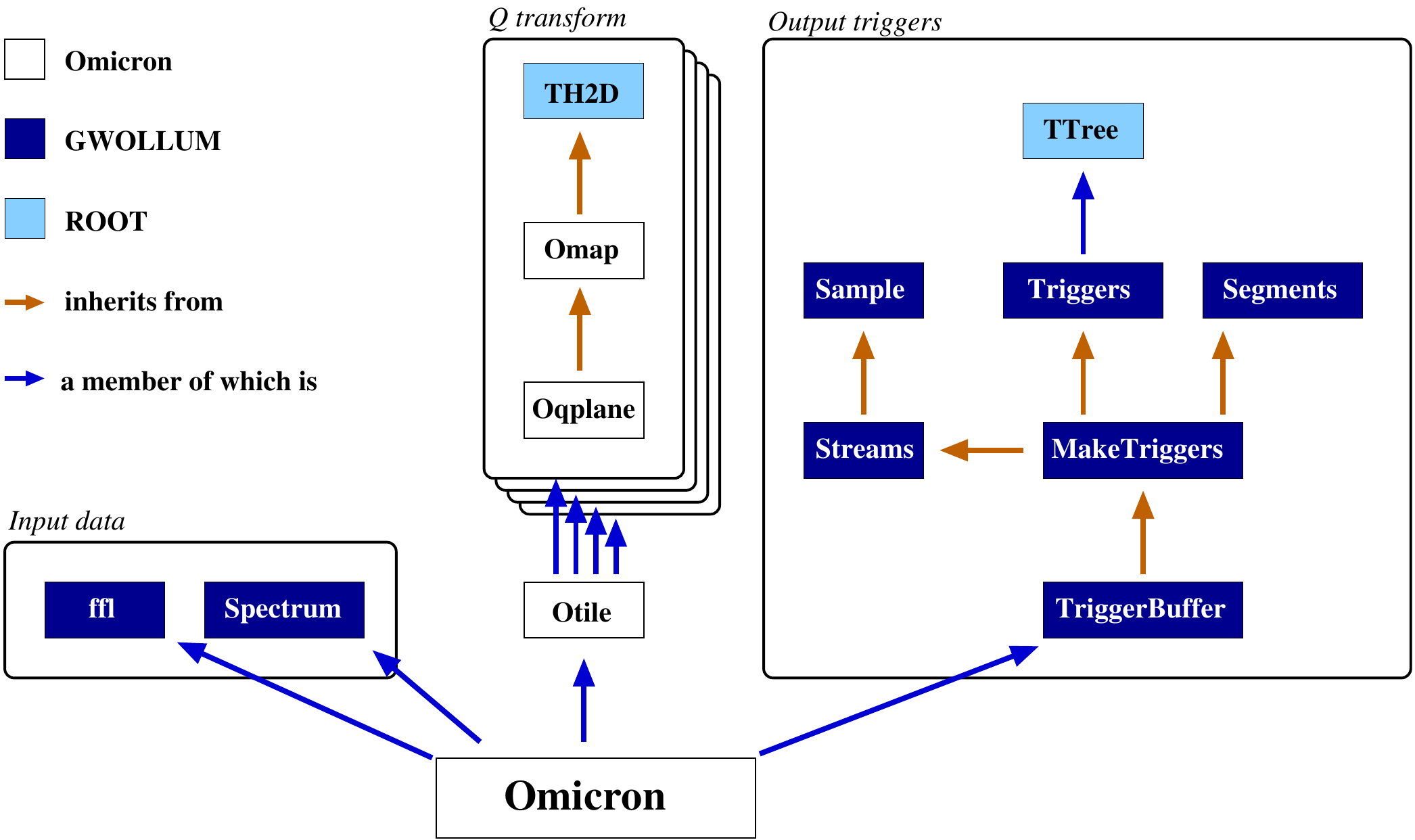}
  \caption{Omicron class organization. See Sec.~\ref{sec:architecture} for a detailed description.}
  \label{fig:classes}
\end{figure*}

\begin{figure*}
  \center
  \includegraphics[width=6.7cm]{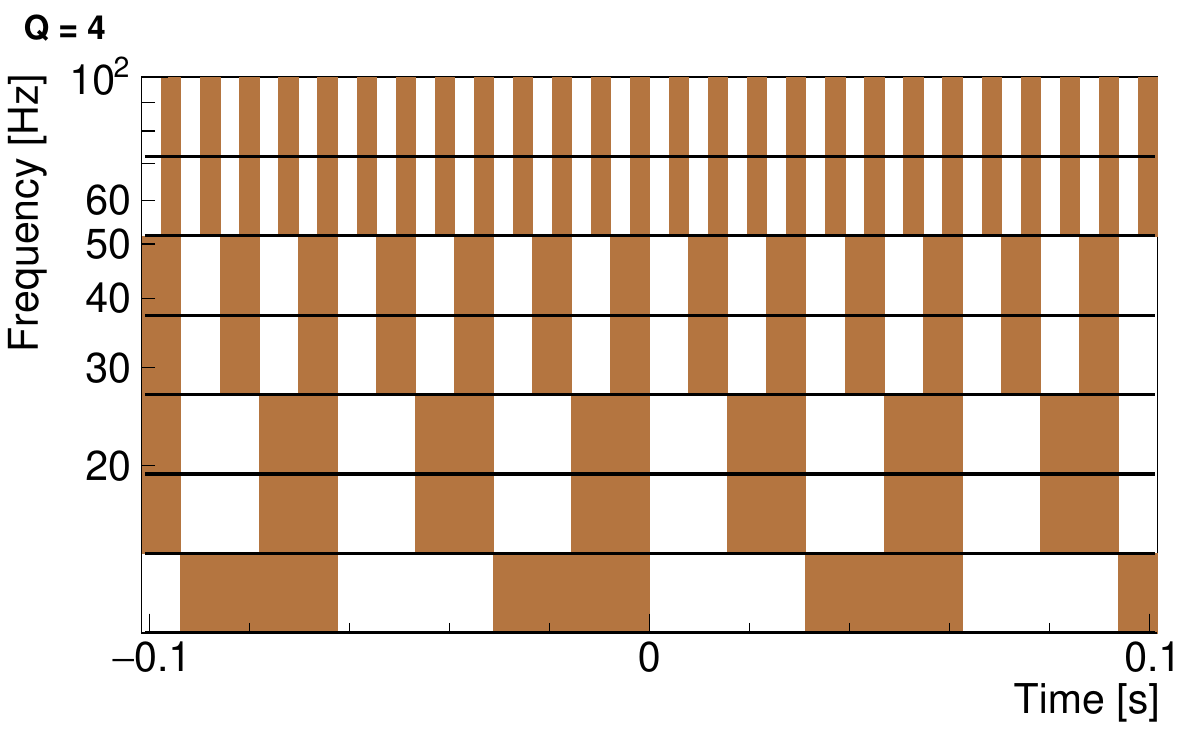}
  \includegraphics[width=6.7cm]{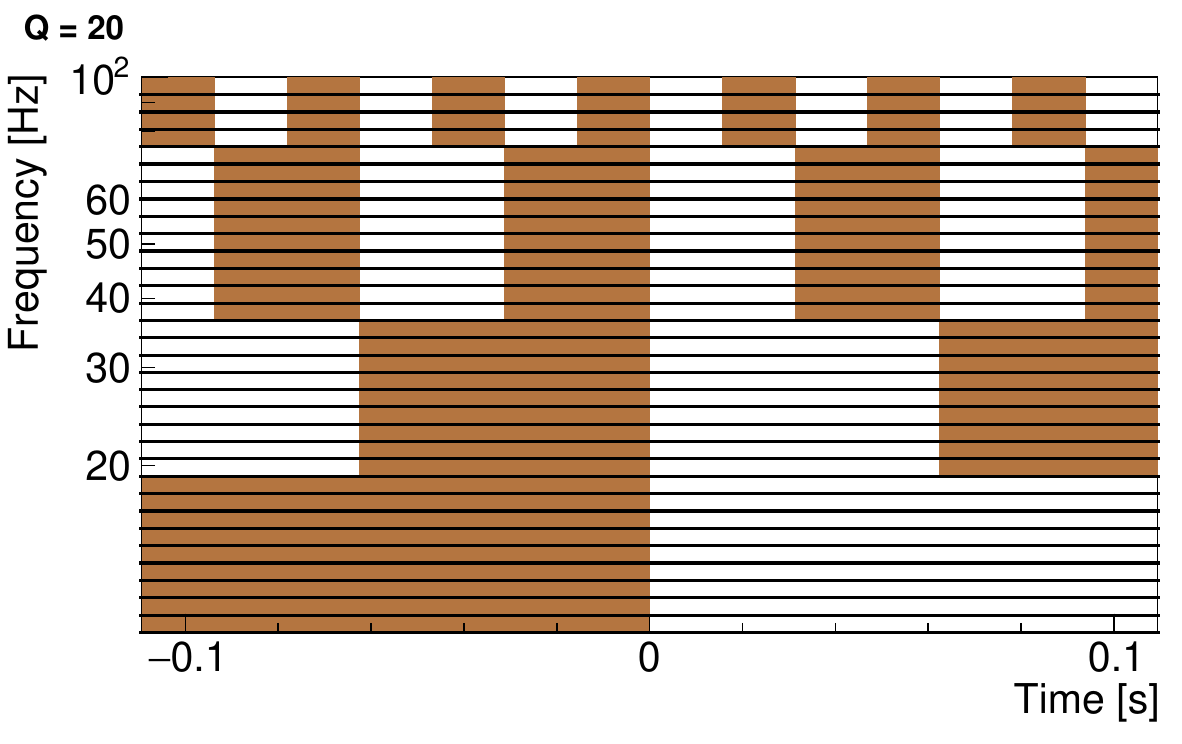}
  \caption{Example of time--frequency planes for two values of $Q$: $Q=4$ (left) and $Q=20$ (right). Each $Q$ plane is tiled (brown and white rectangles) along frequency rows delimited by the horizontal black lines.}
  \label{fig:qplanes}
\end{figure*}

The Omicron code is designed with a modular approach so that new features can easily be added. C++ objects are created to achieve specific analysis tasks, and are orchestrated by a master object from the \texttt{Omicron} class. These objects can be divided in three categories as represented in Fig.~\ref{fig:classes}. Input data are managed by a so-called \texttt{ffl} object which is used to read frame files and extract discrete time series. The time series is first conditioned using various signal analysis filters and windows. Additionally, a \texttt{Spectrum} object is created to dynamically measure the average noise spectrum used to whiten the data.
The $Q$ transform is performed by a class called \texttt{Otile}. This class manages a collection of time--frequency planes called $Q$ planes. A $Q$ plane is represented by a \texttt{Omap} object which can be seen as a two-dimensional histogram~\footnote{Histograms are managed by \texttt{TH2D}~\cite{root-th2d} objects from ROOT.} with bins shaped to describe the time--frequency tiling introduced above. Figure~\ref{fig:qplanes} displays two $Q$ planes, illustrating the multi-resolution analysis performed by Omicron. In virtue of the uncertainty principle, the time resolution increases (while the frequency resolution decreases) when the $Q$ value decreases. The $Q$ plane bins are filled with the tile signal-to-noise ratio estimated with Eq.~\ref{eq:snr} to produce characteristic spectrograms.

Finally, power excesses as measured by Omicron can be saved to disk. Tiles with a signal-to-noise ratio above a given threshold are collected and clustered over time. Resulting events are called triggers and are given parameters, e.g. time, frequency, $Q$, and signal-to-noise ratio, given by the tile with the highest signal-to-noise ratio in the cluster. Tiles and triggers are operated by multiple GWOLLUM classes. Tiles are represented by an object from the \texttt{Triggers} class which inherits from the ROOT \texttt{TTree}~\cite{root-ttree} class developed to manage Ntuples. The \texttt{Triggers} object is supervised by the \texttt{MakeTriggers} class which is designed to save the tiles to disk along with auxiliary data like processing meta-data and analyzed time segments.

\subsection{Software Functionalities}\label{sec:features}
The \texttt{Omicron} class offers many methods for users to conduct their own $Q$ transform analysis. The Omicron package also comes with an out-of-the-box user program which can be run at the command line:
\begin{verbatim}
# run an omicron analysis over one single time segment:
omicron 1234567890 1234568890 ./my_parameters.txt

# run an omicron analysis over a list of time segments:
omicron ./my_segments.txt ./my_parameters.txt
\end{verbatim}
where the Omicron analysis is performed over either one single time segment defined by two GPS times or many GPS time segments listed in a text file (\texttt{./my\_segments.txt}). The analysis parameters are listed in a text file (\texttt{./my\_parameters.txt}) using specific tags and keywords. The input data files must be listed in another text file~\cite{FrameL,LALSUITE} the path of which is specified in the parameter file. One important parameter is the list of output products the Omicron analysis is to generate. Triggers can be saved to disk as described in Sec.~\ref{sec:architecture}. In addition, graphical plots can be generated: spectrograms (single or combined $Q$ planes), time series before and after whitening (including audio tracks), average noise spectra (before and after whitening) and a html report.

\section{Illustrative Examples}\label{sec:examples}
\begin{figure*}
  \center
  \epsfig{width=14cm, file=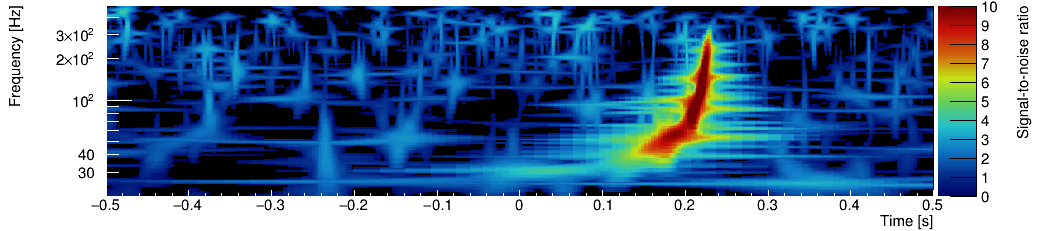} \\
  \epsfig{width=14cm, file=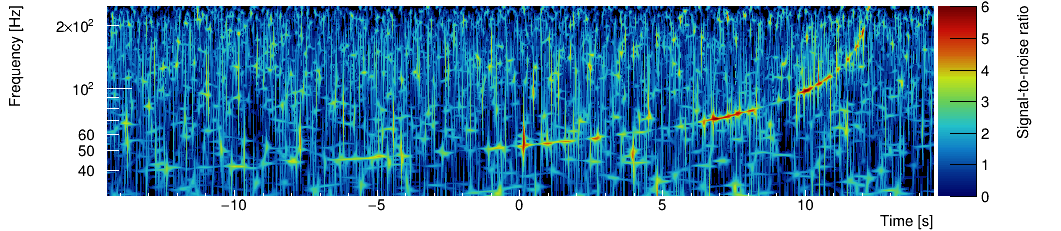}
  \caption{Top: Omicron spectrogram of LIGO--Hanford detector's data around the time of GW150914. The whitened data is projected in multiple time--frequency planes characterized by a constant $Q$ value and the signal-to-noise ratio is measured for each tile. In this representation, all $Q$ planes are stacked up and combined into one;  the tile with the highest signal-to-noise ratio is displayed on top. Bottom: Omicron spectrogram of LIGO--Livingston detector's data around the time of GW170817, using data after glitch subtraction.}
  \label{fig:spectro}
\end{figure*}

The upper plot in Fig.~\ref{fig:spectro} presents the Omicron spectrogram of GW150914 as detected by the LIGO--Hanford interferometer. The frequency evolution of the signal is typical of gravitational waves produced by a binary system of stellar black holes. The different $Q$ planes are stacked up and combined in a single picture, which explains why the power deposits appear star-shaped.

\begin{figure*}
  \center
  \epsfig{width=12cm, file=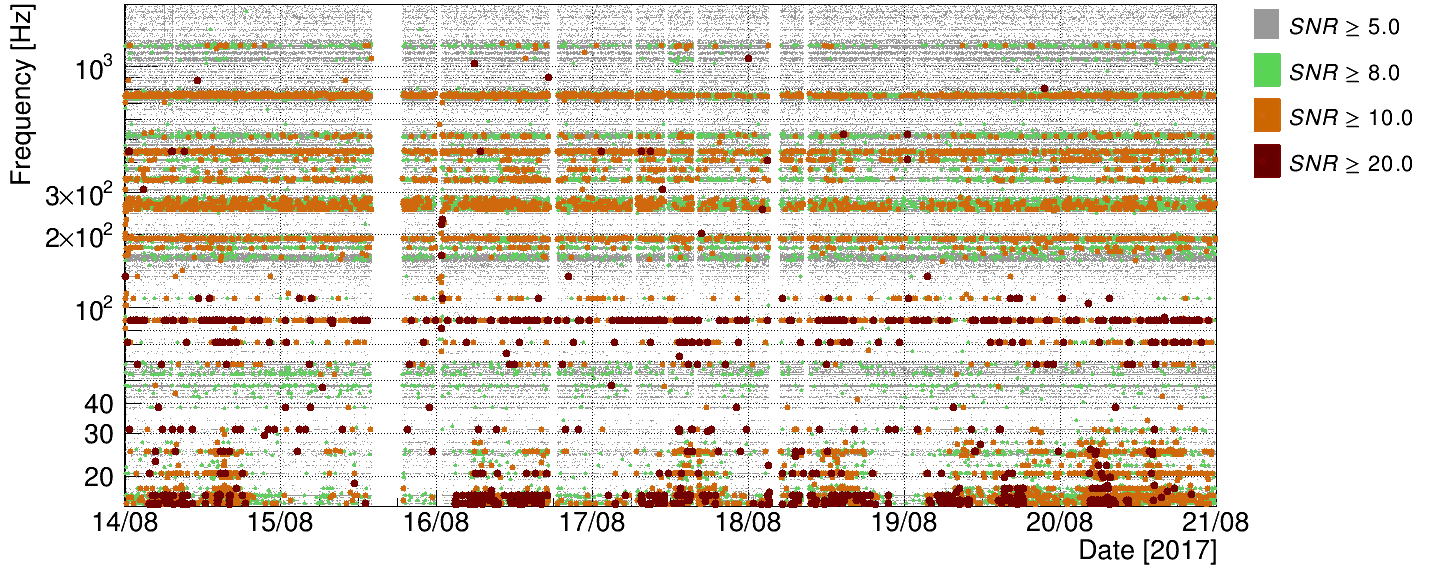}
  \caption{Time--frequency distribution of glitches as detected by Omicron in the Virgo output data over one week in 2017. The glitch signal-to-noise ratio (SNR) is indicated by a color scale. The Omicron processing is interrupted (blank bands) when the detector is not running in nominal conditions.}
  \label{fig:glitchgram}
\end{figure*}

After thresholding the signal-to-noise ratio, tiles are saved in ROOT files. They are typically used to study transient noise in gravitational-wave detectors. For example, in Fig.~\ref{fig:glitchgram}, Omicron triggers generated from Virgo data recorded in August 2017 are distributed in the time--frequency plane. This representation clearly exhibits families of glitches localized in frequency and/or time. This plot is often considered as a starting point for glitch investigations~\cite{verkindt:2019,vim,summary}.

\section{Impact}\label{sec:impact}
The Omicron software is the primary trigger generator used by the LIGO Scientific Collaboration and Virgo collaboration to study and characterize transient noise in the detectors~\cite{McIver:2019hqm}. It is also widely used to vet gravitational-wave detections~\cite{LIGOScientific:2018mvr}. Specific applications~\cite{pyomicron} have been developed and deployed to perform a continuous Omicron analysis over thousands of LIGO and Virgo channels with a low latency. The triggers are saved to disk a few seconds after detector's data are recorded and are available to third-party applications to study transient noise.

For example, algorithms~\cite{Isogai:2010zz,Smith:2011an} were developed to systematically study correlations between glitches measured in the detector's main gravitational-wave channel and triggers detected in auxiliary channels. These analyses typically identify auxiliary channels witnessing a noise disturbance and lead to a better understanding of the noise coupling mechanisms. Triggers generated by Omicron can also be classified into categories often sourced by the same noise. Glitches can be categorized simply using the trigger parameters estimated by Omicron, e.g. the frequency or the signal-to-noise ratio. For example, the low-frequency and high-SNR triggers, below 30~Hz, visible in Fig.~\ref{fig:glitchgram} are the result of seismic noise originating from human activity. Indeed, they are found to coincide with times of working hours and many triggers are coincident in time in auxiliary seismic channels. More advanced methods are also used to classify glitches detected by Omicron based on a multi-variate approach~\cite{Kapadia:2017fhb} or machine-learning techniques~\cite{Biswas:2013wfa,Zevin:2016qwy,Coughlin:2019ref,Mukund:2016thr}. The result of these transient noise investigations can lead to the definition of data quality flags. With this information, LIGO--Virgo gravitational-wave searches can ignore flagged time segments which are contaminated by glitches, hence improving the sensitivity to real transient gravitational-wave events~\cite{TheLIGOScientific:2016zmo,LIGOScientific:2018mvr}.

After the discovery of GW150914, it took several months to vet the gravitational-wave candidate and to certify its astrophysical origin~\cite{TheLIGOScientific:2016zmo}. The Omicron software played an important part to check many aspects related to transient noise. The data surrounding the event was scrutinized and every glitch was carefully inspected. Triggers in auxiliary channels were examined and statistical studies were conducted to discard correlations between GW150914 and transient noise in auxiliary channels.

Two years later, another major event, GW170817, was detected~\cite{TheLIGOScientific:2017qsa}. For the first time, an electromagnetic counterpart~\cite{Ajello:2018mgd} was associated to the source: a binary system of neutron stars. At the time of the event, a loud glitch contaminated the data of one of the LIGO detectors. As a result, the gravitational-wave signal could not be found in coincidence in multiple detectors by online search pipelines. Shortly after the detection alert, the glitch was manually excised and Omicron was used to visualize the signal in a spectrogram. Having the spectrogram of GW170817 for both LIGO detectors provided one more convincing element to guide the decision of releasing the event to astronomers in charge of conducting the follow-up campaign~\cite{GBM:2017lvd}. The bottom plot in Fig.~\ref{fig:spectro} shows the Omicron spectrogram of GW170817 using the final data set where the glitch was subtracted.

After the first discoveries, the process of vetting gravitational-wave candidates was automatized to cope with the increasing rate of detections. When a candidate is identified by online search pipelines, a series of data quality tests is automatically triggered. An Omicron analysis of all LIGO and Virgo channels with a sampling rate above 1~Hz ($\mathcal{O}(1000)$ per detector) is performed to characterize the transient noise at the time of the event: spectrograms of the strain data and auxiliary channels are generated, trigger distributions are plotted, and correlation analyses with auxiliary channels are conducted. The results of these analyses can be reviewed by the scientists and the public alert can be retracted if the data is found to be of bad quality.


\section{Conclusions}\label{sec:conclusion}
The Omicron software occupies a central position in the characterization of transient noise associated to the validation of gravitational-wave events. Omicron combines a high processing speed, high detection efficiency and high resolution to reconstruct parameters of transient noise. Thousands of channels from the LIGO, Virgo and KAGRA~\cite{Akutsu:2017kpk,Akutsu:2019rba} detectors are continuously processed by Omicron. Resulting triggers offer an invaluable starting point for noise investigations. Over the years, Omicron has served the gravitational-wave community at many levels: it helped to mitigate noise issues in the detectors, it was used to produce data quality flags improving the sensitivity of searches, and it characterized the data quality around the time of gravitational-wave events.


\section*{Acknowledgments}
Andrew Lundgren is supported by STFC grant ST/N000668/1. Duncan Macleod is supported by the European Union Horizon 2020 Framework Programme under grant agreement no. 663380-CU073.
This research has made use of data, software and/or web tools obtained from the Gravitational Wave Open Science Center (https://www.gw-openscience.org/ ), a service of LIGO Laboratory, the LIGO Scientific Collaboration and the Virgo Collaboration. LIGO Laboratory and Advanced LIGO are funded by the United States National Science Foundation (NSF) as well as the Science and Technology Facilities Council (STFC) of the United Kingdom, the Max-Planck-Society (MPS), and the State of Niedersachsen/Germany for support of the construction of Advanced LIGO and construction and operation of the GEO600 detector. Additional support for Advanced LIGO was provided by the Australian Research Council. Virgo is funded, through the European Gravitational Observatory (EGO), by the French Centre National de Recherche Scientifique (CNRS), the Italian Istituto Nazionale della Fisica Nucleare (INFN) and the Dutch Nikhef, with contributions by institutions from Belgium, Germany, Greece, Hungary, Ireland, Japan, Monaco, Poland, Portugal, Spain.

\section*{Conflict of Interest}
No conflict of interest exists:
We wish to confirm that there are no known conflicts of interest associated with this publication and there has been no significant financial support for this work that could have influenced its outcome.

\bibliographystyle{elsarticle-num} 
\bibliography{references}

\end{document}